\documentclass[sigconf]{acmart}
\AtBeginDocument{%
  \providecommand\BibTeX{{%
    \normalfont B\kern-0.5em{\scshape i\kern-0.25em b}\kern-0.8em\TeX}}}

\setcopyright{acmcopyright}
\copyrightyear{2023}
\acmYear{2023}
\setcopyright{acmlicensed}\acmConference[KDD '23]{Proceedings of the 29th ACM SIGKDD Conference on Knowledge Discovery and Data Mining}{August 6--10, 2023}{Long Beach, CA, USA}
\acmBooktitle{Proceedings of the 29th ACM SIGKDD Conference on Knowledge Discovery and Data Mining (KDD '23), August 6--10, 2023, Long Beach, CA, USA}
\acmPrice{15.00}
\acmDOI{10.1145/3580305.3599788}
\acmISBN{979-8-4007-0103-0/23/08}

\usepackage{subfigure}
\usepackage{multirow}
\usepackage{bbding}
\usepackage{enumitem}
\usepackage{colortbl}
\usepackage{threeparttable}

\newcommand{\poscl}{\cellcolor{blue!15}}
\newcommand{\negcl}{\cellcolor{gray!18}}
\begin{document}

% \title{Selective Data Replay} 
% \title{How to Predict CVR Accurately in Sales Promotions? A Novel CVR Prediction Framework to Handle Distribution Saltation}

% \title{How to Resolve CVR Fluctuation during Sales Promotion? \\ A Novel Data Reuse Approach for CVR Estimation} 
% \title{How to Address CVR Fluctuation during Sales Promotion? \\ A Novel CVR Estimation Approach via Data Reuse} 
\title{Capturing Conversion Rate Fluctuation during Sales Promotions: \\ A Novel Historical Data Reuse Approach} 
% \title{\chan{Calibrated} Conversion Rate Estimation during Sales Promotion} 
\renewcommand{\shorttitle}{A Novel Historical Data Reuse Approach for Conversion Rate Estimation during Sales Promotions}

\author{Zhangming Chan, Yu Zhang, Shuguang Han*\authornote{Shuguang Han and Baolin Liu are the corresponding authors.}, Yong Bai\textsuperscript{\small{$\dagger$}}, Xiang-Rong Sheng, \\ Siyuan Lou, Jiacen Hu\textsuperscript{\small{$\ddagger$}}, Baolin Liu\textsuperscript{\small{$\ddagger$}}, Yuning Jiang, Jian Xu, Bo Zheng} 
\affiliation{
    \institution{Alibaba Group \textsuperscript{\small{\quad $\dagger$ }}Nanjing University \textsuperscript{\small{\quad $\ddagger$ }}University of Science and Technology Beijing} 
    \city{Beijing, Hangzhou, Nanjing} 
    \country{People's Republic of China} 
}
\email{{zhangming.czm,xieyuan.zy,shuguang.sh,xiangrong.sxr,lousiyuan.lsy,mengzhu.jyn}@alibaba-inc.com} 
\email{baiy@smail.nju.edu.cn, {hujiacen,liubaolin}@ustb.edu.cn, {xiyu.xj,bozheng}@alibaba-inc.com}

\renewcommand{\authors}{Zhangming Chan, Yu Zhang, Shuguang Han, Yong Bai, Xiang-Rong Sheng, Siyuan Lou, Jiacen Hu, Baolin Liu, Yuning Jiang, Jian Xu, Bo Zheng}
\renewcommand{\shortauthors}{Zhangming Chan et al.}

\begin{abstract}
  Conversion rate (CVR) prediction is one of the core components in online recommender systems, and various approaches have been proposed to obtain accurate and well-calibrated CVR estimation. However, we observe that a well-trained CVR prediction model often performs sub-optimally during sales promotions. This can be largely ascribed to the problem of the data distribution shift, in which the conventional methods no longer work. To this end, we seek to develop alternative modeling techniques for CVR prediction. Observing similar purchase patterns across different promotions, we propose reusing the historical promotion data to capture the promotional conversion patterns. Herein, we propose a novel \textbf{H}istorical \textbf{D}ata \textbf{R}euse (\textbf{HDR}) approach that first retrieves historically similar promotion data and then fine-tunes the CVR prediction model with the acquired data for better adaptation to the promotion mode. HDR consists of three components: an automated data retrieval module that seeks similar data from historical promotions, a distribution shift correction module that re-weights the retrieved data for better aligning with the target promotion, and a TransBlock module that quickly fine-tunes the original model for better adaptation to the promotion mode. Experiments conducted with real-world data demonstrate the effectiveness of HDR, as it improves both ranking and calibration metrics to a large extent. HDR has also been deployed on the display advertising system in Alibaba, bringing a lift of $9\%$ RPM and $16\%$ CVR during Double 11 Sales in 2022. 
\end{abstract} 





\maketitle

\section{Introduction}
\label{sec:intro}

As one of the largest e-commerce platforms in the world, Taobao provides an online shopping service for hundreds of millions of users on a daily basis. To facilitate an effective matching of user interest with millions of online products, we spend significant efforts developing deep machine learning models on Taobao.
Among them, the click-through rate (CTR) prediction model~\cite{sheng2021model,zhang2022keep,sheng2022joint,bian2022can,wu2022adversarial} and the conversion rate (CVR) prediction model are the two most common ones. In this study, we are interested in the CVR prediction problem. Researchers and practitioners have proposed a variety of approaches for obtaining accurate and well-calibrated estimation of CVR~\cite{ma2018entire,wen2020entire,wen2021hierarchically,gu2021real,chen2022asymptotically,xu2022ukd,huangfu2022multi,yang2022generalized}. For instance, to tackle the problem of data sparsity, \citet{wen2021hierarchically,wen2020entire} and Yang et al.~\cite{yang2022generalized} utilized a series of post-click behaviors for supplementing the rare conversion event, which resulted in an improved performance on CVR prediction. Other researchers dealt with the selection bias problem in CVR prediction and proposed the entire space estimation method by jointly learning with multiple behavior prediction tasks~\cite{ma2018entire, xu2022ukd}. 

\begin{figure}[t]
    \centering
    \includegraphics[width=1.0\columnwidth]{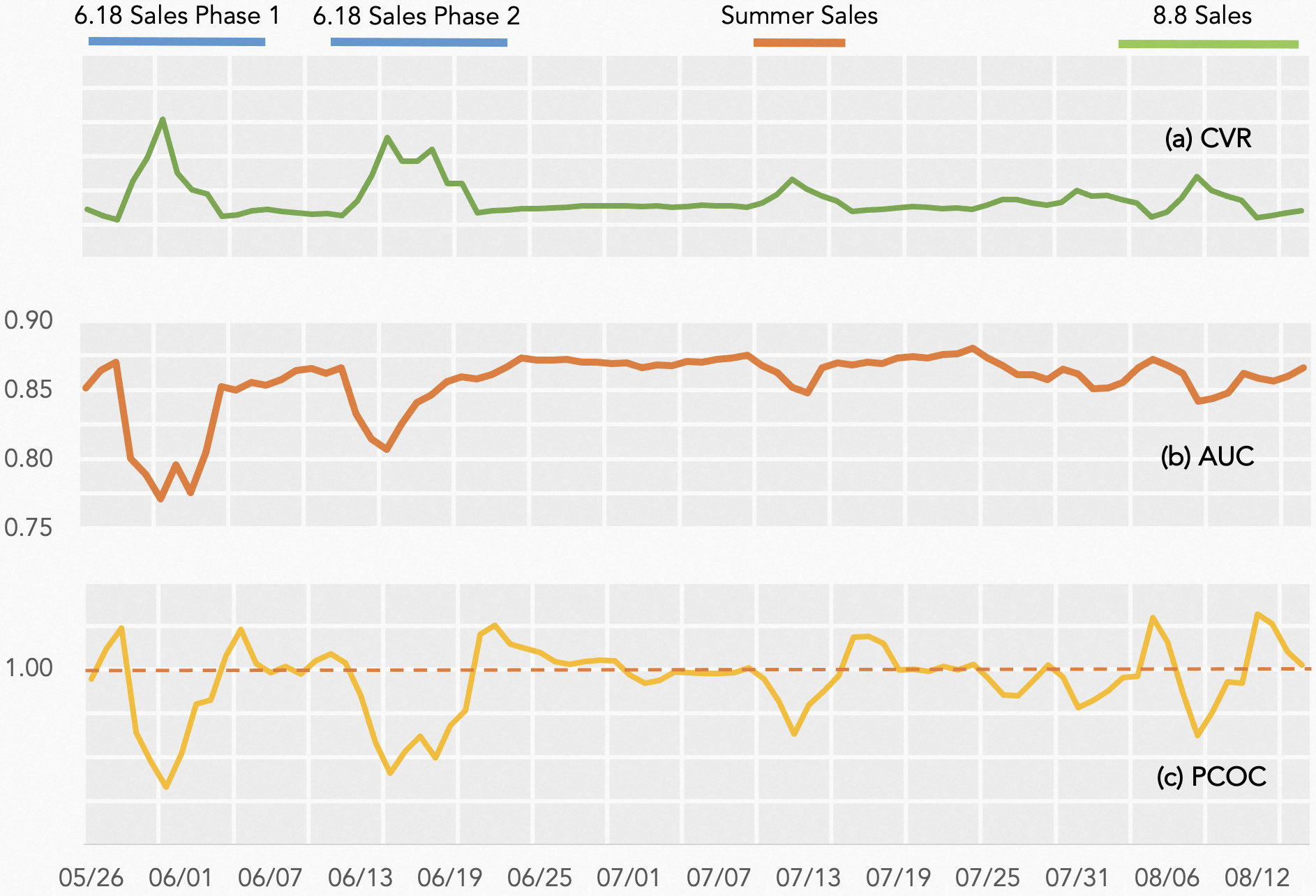} 
    % \caption{An illustration of the actual CVR and AUC/PCOC performance for the production CVR prediction model. Here, X-axis represents the date, and Y-axis denotes the corresponding value of each metric. A PCOC value of 1.0 indicates perfect calibration. For the purpose of data anonymity, we keep a minimal number of labels on the Y-axis for CVR and PCOC. It is clear that the CVR fluctuates during sales promotions, and AUC/PCOC decreases accordingly.} 
    \caption{An illustration of the actual CVR and AUC/PCOC performance for the production CVR prediction model. To ensure data anonymity, we limit the number of labels on the Y-axis for CVR and PCOC to a minimum.} 
    \label{fig:metrics-in-promotions}  
\end{figure}

Despite the previous advancement, we still find that our production CVR prediction model performs sub-optimally during \textbf{sales promotions}. Sales promotion is a popular marketing strategy in e-commerce platforms that stimulate consumers' demand by offering a significant discount within limited time. The Black Friday Sales on Amazon and the Double 11 Sales on Taobao are the two well-known online promotion festivals worldwide. It is worth noting that sales promotion is frequently held on e-commerce platforms. For example, there are 33 promotions in Taobao 2022, taking about half of the time in a year. According to Figure~\ref{fig:metrics-in-promotions}(b) and Figure~\ref{fig:metrics-in-promotions}(c), there is a deteriorated performance of the production model during a sales promotion. Both the AUC
% \footnote{\textbf{A}rea \textbf{U}nder receiver operating characteristic \textbf{C}urve, which is commonly adopted to measure the ranking performance.} 
and PCOC\footnote{\textbf{P}redicted \textbf{C}VR \textbf{O}ver the actual \textbf{C}VR, $\text{PCOC}=\frac{\sum \text{predicted CVR}}{\sum \text{actual CVR}}$ is generally applied for evaluating the calibration ability, with 1.0 denoting a perfect calibration.} metrics drop significantly. For a concrete example, during the 6.18 Sales, the online AUC falls as much as 10\%, and the value of PCOC is only 1/3. This seriously affects the effectiveness of the production system.

Such a phenomenon can be primarily ascribed to the rapid change of data distribution, as we observe a substantial fluctuation of the actual CVR during sales promotion, as shown in Figure~\ref{fig:metrics-in-promotions}(a). Therefore, conventional CVR prediction models with i.i.d. (independent and identically distributed) assumption on the training and test data would certainly fail during the promotion. To resolve this problem, one may improve the model freshness so that the prediction model can be quickly adapted to the \textit{promotion mode}. However, the delayed feedback nature of the conversion event further complicates the utilization of recent data.

Unlike click behaviors, feedback labels for the conversion event cannot be collected within a short time as conversion may happen in days or even weeks afterward. This introduces a dilemma -- to employ recent data with incomplete labels or to wait for more accurate labels. By assuming a stable delayed conversion pattern (e.g., the delay time for the final conversion remains steady), a delayed feedback model resolves this problem by utilizing the relationship between the ultimate conversion and the conversion time to correct the incomplete real-time labels. 
However, such an assumption does not hold during the sales promotion since there is a rapid fluctuation of conversion behaviors at this time, which calls for the development of alternative modeling techniques.

In spite of the discrepancy between non-promotional and promotional conversions, the conversion patterns in different sales promotions remain to be similar (see more details in Figure~\ref{fig:conversion-pattern} and Table~\ref{tab:label-shift}). This motivates us to fine-tune our model with the historical data from similar promotions. Since the past promotions have already finished and the complete conversion labels are now easily accessible, the historical data no longer encounter the delayed feedback problem. To this end, we propose a novel \textbf{H}istorical \textbf{D}ata \textbf{R}euse (\textbf{HDR}) approach with the first stage concentrating on retrieving historically similar promotional data, and the second stage focusing on model fine-tuning with the retrieved data. 

Specifically, we first represent each day with a vector of designed features. Then, for a target promotion, we search for the most similar promotion in history with the nearest neighbor algorithm. After that, the retrieved data is applied for model fine-tuning so that the model can quickly adapt to the promotion mode. To further address the disparity of conversion behaviors between the retrieved promotion and the target promotion, we design a distribution shift correction method to further refine the fine-tuning mechanism. Concretely, we first estimate an importance weight to measure the discrepancy of different promotions. Then, we utilize the obtained importance weight to revise the fine-tuning loss following the Importance-Weighted Empirical Risk Minimization framework~\cite{shimodaira2000improving}.

In practice, the model fine-tuning with the historical data often encounters the one-epoch problem~\cite{zhang2022towards}, i.e., the model will be overfitted when the same data has been trained twice. For this reason, we design a TransBlock module (see Figure~\ref{fig:transblock}) to isolate the fine-tuning parameters from the main model and employ two separate training configurations. To capture different conversion patterns, TransBlock explicitly models the transition between promotional conversions and non-promotional conversions. In this way, the main model trained from the non-promotional conversion data will be less impacted, while the fine-tuning module helps the production model quickly adapt to the promotion mode.

The main contributions of our work are summarized as follows:
\begin{itemize}[leftmargin=*]
\item With regard to the deteriorated performance of the production CVR prediction model during sales promotions, we propose a novel Historical Data Reuse (HDR) algorithm, aiming to learn common patterns from different historical promotions.

\item To address the overfitting problem from a direct fine-tuning with the historical data, we design a TransBlock module that separates the fine-tuning parameters from the main model and explicitly models the transitions between the non-promotional conversion to the promotional conversion.

\item HDR has been successfully deployed in the display advertising system of Alibaba, bringing a substantial lift of core production metrics (RPM+9\% and CVR+16\%) during Double 11 Sales in 2022. 
\end{itemize}
 
\section{Related Work} 
\label{sec:related}
CVR prediction during sales promotion is a complicated problem that suffers from the distribution shift and delayed conversions. To provide a comprehensive overview of relevant techniques, we introduce a variety of related works, including conversion rate prediction models and distribution shift correction algorithms. 

\subsection{Conversion Rate Prediction} 
CVR prediction is essential for many industrial machine-learning systems, including search, recommendation, and online advertising. A variety of approaches have been proposed to deal with the accurate estimation of conversion rate~\cite{wen2021hierarchically,wen2020entire,ma2018entire,xu2022ukd,wang2020delayed,gu2021real,chen2022asymptotically,huangfu2022multi,yang2022generalized}. In practice, due to the inherent similarity in training sample construction, CVR model often adopts a resembling modeling architecture as the CTR model. Previous studies have also developed a set of novel solutions in considering the unique characteristics of CVR prediction. For instance, to address the data sparsity issue, researchers have used a series of post-click behaviors to supplement the rare conversion event~\cite{wen2021hierarchically,wen2020entire,yang2022generalized}. Other researchers focused on dealing with the problem of selection bias in CVR estimation and proposed the entire space estimation method by jointly learning with several other user behavior prediction tasks~\cite{ma2018entire, xu2022ukd}. 

\subsection{Delayed Feedback Modeling} 
Since the conversion events may happen long after clicks, the collection of actual conversion labels may delay by hours or even days. The importance of delayed feedback modeling is first emphasized in the pioneering work DFM by~\citet{chapelle2014modeling}. DFM introduces an additional model that captures the expected conversion delay, assuming the delay distribution follows the exponential distribution. 
% \citet{ktena2019addressing} first studied the delayed feedback problem under the online training setting. They treated all data samples as negatives in the beginning. Once the positive engagement occurs, the sample will be duplicated with a positive label and trained for the second time. Importance sampling is adopted to re-weight training samples to further learn the model from the bias distribution. 
\citet{ktena2019addressing} first studied the delayed feedback problem under the online training setting. They treated all data samples as negatives in the beginning. Once the positive engagement occurs, the sample will be duplicated with a positive label and trained for the second time. Importance sampling is adopted to re-weight training samples to further learn the model from the bias distribution. However, the resulting performance can be less satisfied due to the inaccuracy of weight computation. 
Therefore, \citet{yang2021capturing} proposed to wait longer to obtain more conversions. In the meantime, the delayed positive samples were also duplicated in the training pipeline. \citet{gu2021real} and \citet{chen2022asymptotically} argued that the duplicated positive samples change the training distribution, which leads to sub-optimal performance. Hence, they proposed duplicating real negative samples and refining the importance weights to maintain consistency between training and testing distribution.

The above approaches often assume a stable delayed conversion pattern, e.g., the delay distribution for the conversion remains the same between training and test data. This presumption may hold in non-promotional data. However, as mentioned in Section~\ref{sec:intro}, such a presumption does not hold during the sales promotion. 

\subsection{Distribution Shift} 
The fluctuation of actual CVR during sales promotion is mainly due to the problem of the rapid distribution shift. Distribution shift states that there is an inconsistency between the training distribution $p(x,y)$ and the testing distribution $q(x,y)$. Previous research has studied this problem from two perspectives: covariate shift~\cite{shimodaira2000improving,bickel2009discriminative,wang2020tent,chen2022contrastive,choi2022improving,niu2022efficient,liu2021ttt} and label shift~\cite{scholkopf2012causal,2018Detecting,2019Regularized,alexandari2020maximum,wu2021online,bai2022adapting}.

Covariate shift assumes that $p(y|x)$ is constant and $p(x)$ changes between training and testing. Under this presumption, \citet{shimodaira2000improving} proposed an Importance-Weighted Empirical Risk Minimization (IW-ERM) framework that utilizes importance weights to adjust the training distribution. Similar approaches have also been adopted by other researchers~\cite{bickel2009discriminative,huang2006correcting}. Different from the covariate shift, label shift states that $p(x|y)$ is stable between training and testing distributions. Black Box Shift Estimation (BBSE) is such an algorithm to tackle the problem of label shift with a black-box predictor~\cite{2018Detecting}. In addition, \citet{scholkopf2012causal} proposed to utilize causal inference to solve the label shift problem, and \citet{2019Regularized} developed a Regularized Learning under Label Shift (RLLS) approach that adopted regularization to de-bias the BBSE algorithm. Note that BBSE and RLLS are also the basis of our work.

% which applies a domain-adaptation algorithm to revise label shift. 
% This approach uses the labeled training data and the unlabeled target data to estimate importance weights, then retrain the predictor in BBSE on the weighted training data.  

% \section{Methodology} 
\section{Methodology} 
\label{sec:methodology}

\begin{figure}[t]
    \centering
    \includegraphics[width=\columnwidth]{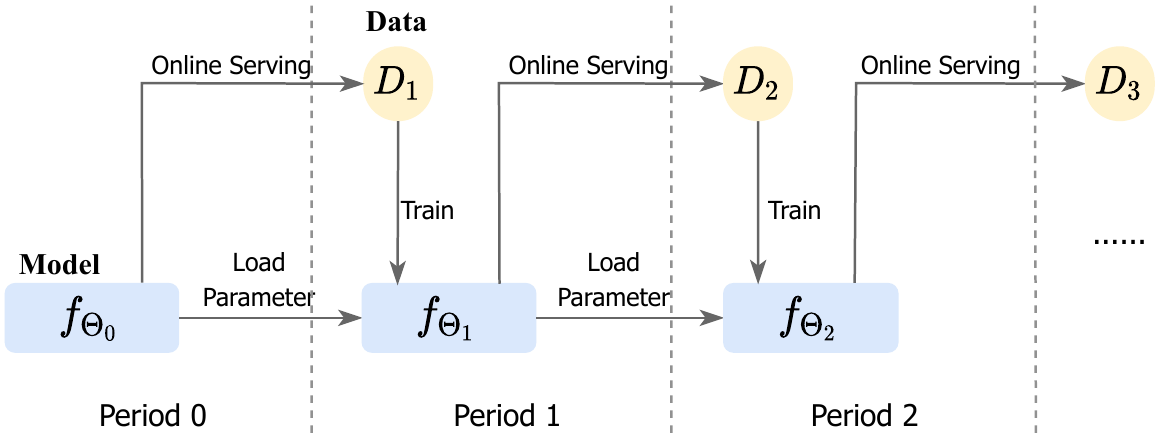} 
    \caption{An illustration of the online learning process for our production CVR model, which is updated on a daily basis.} 
    \label{fig:odl}
    % \vspace{-1em}
\end{figure}

Before delving into the details of our proposed approach, we first provide a formal description of the problem. Let $f_\Theta(\cdot)$ denote the CVR prediction model, and $\Theta$ indicates the trainable parameters. Our production model optimizes $\Theta$ with the training data $\mathcal{A}(x,y)$ in a supervised learning way. After being deployed online, it will be used to serve an incoming new data distribution $\mathcal{B}(x,y)$. In practice, as shown in Figure~\ref{fig:odl}, our production model is trained in an online learning manner~\cite{luo2021Bernoulli,zhang2022keep,bian2022can} and is updated on a daily basis to make a trade-off between delayed conversion and data freshness. 

The validity of the above training paradigm relies on the i.i.d. assumption between $\mathcal{A}$ and $\mathcal{B}$. However, as illustrated by Figure~\ref{fig:metrics-in-promotions}, this hypothesis fails during sales promotions. To deal with the data distribution shift problem, a potential solution is to seek a historical distribution $\mathcal{B'}$ that resembles the incoming promotion data $\mathcal{B}$, and further fine-tune the production model with $\mathcal{B'}$. This can be formalized using Formula~\ref{eq:training-framework}, in which $f_\Theta'$ denotes the fine-tuned model that is meant to adapt the promotion mode. Note that instead of directly training a model with $\mathcal{B}'$ (while discarding $\mathcal{A}$), we adopt the \textbf{training-and-finetuning} approach to guarantee the recent data remain to be included in the model.
\begin{equation}
% \small 
\label{eq:training-framework} 
f_\Theta: Train(\mathcal{A}) \Longrightarrow
f_\Theta': Train(\mathcal{A}) + Finetune(\mathcal{B}')
\end{equation}

The above formulation introduces several important challenges: \textbf{(1)} how to obtain the historical promotion data distribution $\mathcal{B}'$ that is similar to $\mathcal{B}$, \textbf{(2)} how to deal with the discrepancy, if any, between $\mathcal{B}$ and $\mathcal{B}'$, and \textbf{(3)} what are the effective architectures for model fine-tuning over $\mathcal{B}'$. The proposed HDR approach focuses on tackling these three obstacles, and the detailed implementations are provided in the below subsections.

\subsection{Obtaining $\mathcal{B'}$ with Historical Data Retrieval} 
\label{sec:retrieval}
For challenge (1), we design a simple yet effective vector-based approach to retrieve historically similar promotion data distribution $\mathcal{B'}(x,y)$. In our production system, the historical data is processed in the granularity of natural days, we thus regard each day as a candidate distribution for retrieval. 

\begin{figure}[!tbp]
    \centering
    \includegraphics[width=0.9\columnwidth]{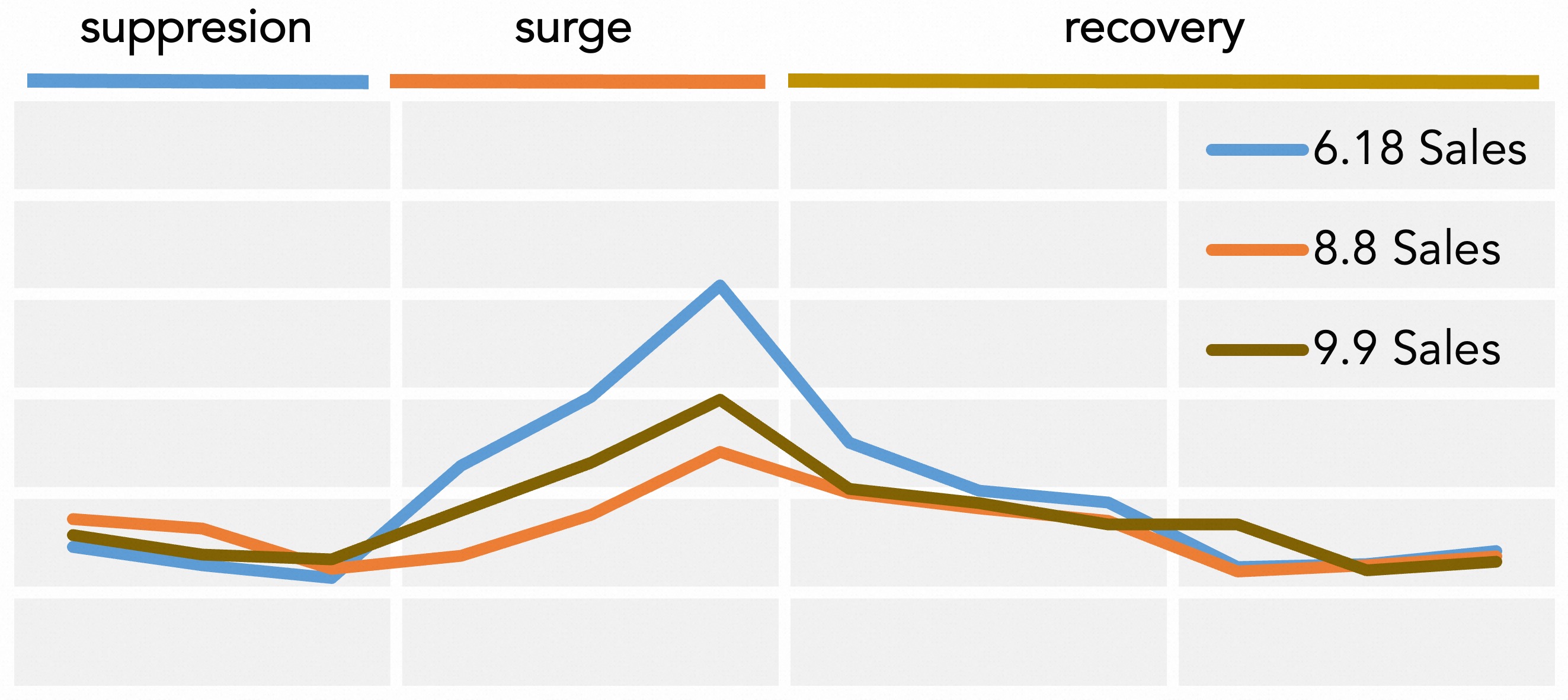} 
    \caption{The change of conversion rate (Y-axis) over time in three different sales promotions. They all exhibit a similar pattern: pre-promotion suppression, on-promotion surging, and post-promotion recovery. For the purpose of data anonymity, we hide the labels on the Y-axis.} 
    \label{fig:conversion-pattern}
    % \vspace{-1em}
\end{figure}

\subsubsection{Vector-Based Representation} 
\label{subsec:hdrr-vector-repsentation}
To facilitate automated data retrieval, we first represent each day with a vector of numerical features. As illustrated by Figure~\ref{fig:conversion-pattern}, the conversion pattern in different sales promotions looks quite similar -- they exhibit a resembling pattern of pre-promotion suppression, on-promotion surging, and post-promotion recovery. By capturing such a trend, we can easily distinguish ordinary time from promotion activity. To this end, we adopt two kinds of \textbf{conversion-related features} from the most recent three days when computing the data similarity. 

\textbf{CVRs from previous days.} Average CVR per day is obviously a good indicator of different conversion patterns. We first adopt a set of CVR metrics from the most recent three days for vector representation. Note that our production system utilizes three days as the attribution window; thereby, a non-conversion within one or two days (incomplete attribution) may end up with a conversion in three days (complete attribution). We thus consider both complete and incomplete attribution of conversions, including the conversion in one day, two days, and three days, for vector representation.

\textbf{Impression ratio of representative categories.} In practice, we find that the purchase of certain product categories such as cosmetics often surges during promotion, and this phenomenon is consistent across different sales promotions. As a consequence, we consider several representative categories of such and utilize the impression ratio of each category as a numerical feature. Here, the impression ratio of a category is computed with the number of impressions for the given category over the total number of impressions. Meanwhile, these categories are selected based on the discrepancy of impression ratio between the promotion and non-promotion periods of time, which include Toddler, House Cleaning, Personal Care, Cosmetics and etc. 

Since the impression ratio does not rely on the collection of user feedback, it can be readily available. Therefore, we further compute the impression ratios with the data from the first ten hours of a target promotion day. These new features result in a performance boost for data retrieval and lead to a further improvement in online performance despite a 10-hour delay for model serving.

\subsubsection{The Retrieval Process} 
With the vector-based representation, we then utilize the nearest neighbor algorithm for retrieving the most similar historical data distribution. Here, cosine distance is adopted as the similarity measure while searching for the nearest neighbors. In our production system, the retrieved top-two days will be kept for later model fine-tuning.

\subsection{Distribution Shift Correction}
\label{sec:dsc}
As mentioned in challenge (2), despite being similar, there is no guarantee of complete equivalence for the two data distributions $\mathcal{B}(x,y)$ and $\mathcal{B'}(x,y)$. Figure~\ref{fig:conversion-pattern} demonstrates that there are still discrepancies in the actual conversion rates for different sales promotions, particularly at the on-promotion surging phase. This calls for the fine-tuning module, as formulated by Equation~\ref{eq:training-framework}, to come out with a resolution for the distribution shift problem.

As a consequence, we redesign the fine-tuning module under the Importance-Weighted Empirical Risk Minimization (IW-ERM) framework~\cite{2000Improving} by minimizing the below empirical risk:
\begin{equation}
% \small 
\begin{aligned}
\label{eq:IWERM} 
\mathcal{L} &= \int \frac{\mathcal{B}(x,y)}{\mathcal{B}'(x,y)} \cdot \ell(x,y) \ \mathrm{d} \mathcal{B}'(x,y)
\end{aligned}
\end{equation}
where $\ell(x,y)$ denotes the standard cross-entropy loss computed with the distribution $\mathcal{B}'(x,y)$. $\frac{\mathcal{B}(x,y)}{\mathcal{B}'(x,y)}$ measures the importance weight that aims to correct the disparity between the two distributions. With Bayes' theorem, this can be further transformed to:
\begin{equation}
% \small 
\begin{aligned}
\label{eq:IWERM-bayes} 
\mathcal{L}
&= \int \frac{\mathcal{B}(x|y)}{\mathcal{B}'(x|y)} \cdot \frac{\mathcal{B}(y)}{\mathcal{B}'(y)} \cdot \ell(x,y) \ \mathrm{d} \mathcal{B}'(x,y)
\end{aligned}
\end{equation}
in which $\mathcal{B}(x|y)$ and $\mathcal{B}'(x|y)$ represent the conditional input distributions, $\mathcal{B}(y)$ and $\mathcal{B}'(y)$ stand for the label distributions. For better understanding, we re-emphasize that $\mathcal{B}(x,y)$ denotes the target promotion data distribution, and $\mathcal{B'}(x,y)$ refers to the data distribution retrieved from historical promotion data. However, since $\mathcal{B}(y)$, $\mathcal{B}(x|y)$ and $\mathcal{B}(x,y)$ are both unavailable at the serving time, we need to make a few presumptions.

\begin{table} 
\renewcommand{\arraystretch}{0.9} 
\centering 
% \small 
\caption{An analysis of the conditional input distribution $p(x|y)$ for 8.8 Sales and 9.9 Sales. We examine the conversion patterns for two groups of users ($\text{U}_1$, $\text{U}_2$) and two product categories ($\text{C}_1$, $\text{C}_2$). The value in each cell denotes the corresponding conditional probability.} 
% \caption{An analysis of the conditional input distribution $p(x|y)$ for 8.8 Sales and 9.9 Sales. We examine the conversion patterns for two groups of users ($\text{U}_1$, $\text{U}_2$) and two product categories ($\text{C}_1$, $\text{C}_2$).} 
\label{tab:label-shift} 
\resizebox{\columnwidth}{!}{
\begin{tabular}{c|cc|cc} 
\toprule 
\multirow{2}{*}{User Group $\times$ Category} & \multicolumn{2}{c}{8.8 Sales} & \multicolumn{2}{|c}{9.9 Sales} \\ \cmidrule(lr){2-5}
& $y=0$ & $y=1$ & $y=0$ & $y=1$ \\ 
\midrule 
$\text{U}_{1}, \text{C}_{1}$ & 0.9597\% & 0.3831\% & 1.1013\% & 0.3595\% \\ 
$\text{U}_{1}, \text{C}_{2}$ & 0.0064\% & 0.0472\% & 0.0065\% & 0.0575\% \\ 
\midrule 
$\text{U}_{2}, \text{C}_{1}$ & 0.3125\% & 0.1738\% & 0.3441\% & 0.1597\% \\ 
$\text{U}_{2}, \text{C}_{2}$ & 0.1769\% & 0.0258\% & 0.1738\% & 0.0288\% \\ 
\bottomrule 
\end{tabular}} 
%  \begin{tablenotes}
%     \footnotesize 
%     \item[*] * The value in each cell denotes the corresponding conditional probability.
% \end{tablenotes}
\end{table}

\subsubsection{Presumption I: $\mathcal{B}(x|y) = \mathcal{B}'(x|y)$} \label{subsec:presumption1}
We first assume the equivalence of the two conditional input distributions $\mathcal{B}(x|y)$ and $\mathcal{B}'(x|y)$, which means that among the purchased (non-purchased) items, the input distributions stay to be similar. It is worth noting that this actually shares the same presumption as the Black Box Shift Estimation (BBSE) algorithm~\cite{2018Detecting}, which is also applied for estimating the target label distribution $\mathcal{B}(y)$ later on.

To seek evidence from the actual data, we analyze the conditional input distribution with two sales promotions. More specifically, we select user group and product category as two representative input features and fill the conditional probabilities $p(\text{U}_i, \text{C}_j|y=0)$ and $p(\text{U}_i, \text{C}_j|y=1)$ in Table~\ref{tab:label-shift}. It is apparent that $p(x|y)$ holds steadily across different sales promotions, lending significant support for the above presumption. With such an assumption, the loss function can be further approximated by:
\begin{equation}
% \small 
\begin{aligned}
\label{eq:loss-after-presumption1} 
\mathcal{L} = \int \frac{\mathcal{B}(y)}{\mathcal{B'}(y)} \cdot \ell(x,y) \ \mathrm{d} \mathcal{B'}(x,y) 
\end{aligned} 
\end{equation}

To be consistent with the model update frequency, the label distributions $\mathcal{B}(y)$ and $\mathcal{B}'(y)$ are estimated in the granularity of a natural day. Here, $\mathcal{B}'(y)$ can be easily computed with the historical data while $\mathcal{B}(y)$, by the nature of delayed feedback, remains unavailable at the serving time. Inspired by ~\citet{2018Detecting}, we develop a distribution estimation approach for obtaining $\mathcal{B}(y)$. More specifically, we first introduce $\hat{y}$ to indicate the model prediction of the input $x$, i.e., $f_{\Theta}(x)$. Since $\hat{y}$ only relies on the data input, with Presumption I, we can easily demonstrate the equivalence of $\mathcal{B}(\hat{y}|y)$ and $\mathcal{B}'(\hat{y}|y)$. In this way, the distribution of model prediction $\mathcal{B}(\hat{y})$ can be expressed with the below Equation~\ref{eq:bbse}. 
% \begin{equation}
% \begin{aligned}
% \label{eq:bbse} 
% \mathcal{B}(\hat{y}) 
% &= \int \mathcal{B}(\hat{y}|y) \cdot \mathcal{B}(y) \ \mathrm{d}y \qquad \textit{(Total Probability)}\\
% &= \int \mathcal{B'}(\hat{y}|y) \cdot \mathcal{B}(y) \ \mathrm{d}y \qquad \textit{(Presumption I)}
% \end{aligned} 
% \end{equation} 
\begin{equation}
% \small 
\begin{aligned}
\label{eq:bbse} 
\mathcal{B}(\hat{y}) 
&= \sum_{y \in \{0, 1\}} \mathcal{B}(\hat{y}|y) \cdot \mathcal{B}(y) \qquad \textit{(Total Probability)}\\
&= \sum_{y \in \{0, 1\}} \mathcal{B'}(\hat{y}|y) \cdot \mathcal{B}(y), \qquad \textit{(Presumption I)}
\end{aligned} 
\end{equation} 

\begin{figure}[t]
    \centering
    \includegraphics[width=\columnwidth]{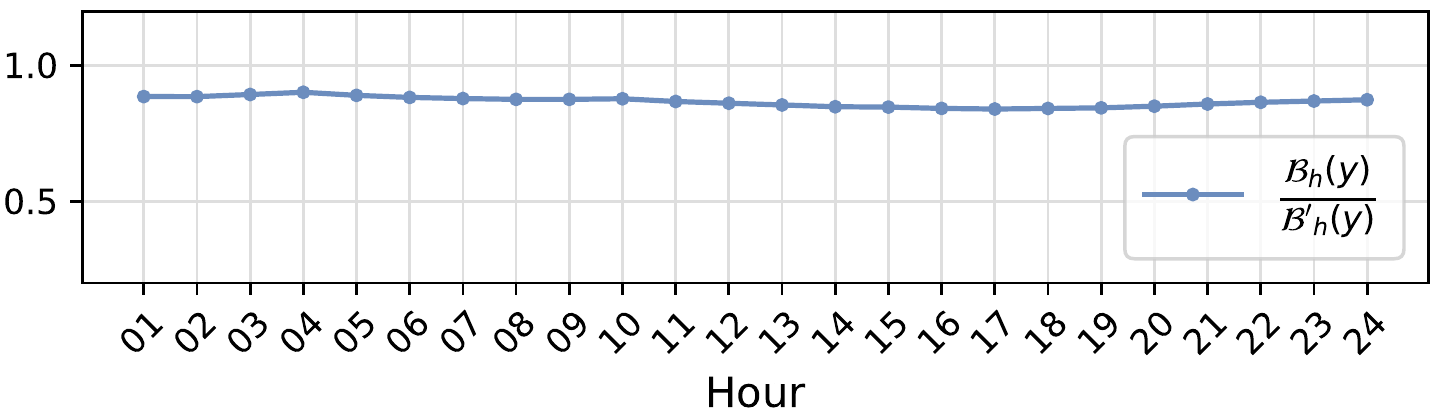} 
    \caption{An illustration of $\mathcal{B}_h(y)/\mathcal{B}'_h(y)$ over different selections of the hour $h$. Here, the largest difference is less than 5\% as the maximum and minimum values are 0.89 and 0.85, respectively. This clearly shows the stability of the ratio.} 
    \label{fig:cvr-ratio}
    \vspace{-1em}
\end{figure}

Once $\mathcal{B}(\hat{y})$ and $\mathcal{B}'(\hat{y}|y)$ are both available, one can solve this Equation and obtain an analytic solution for $\mathcal{B}(y)$. However, $\mathcal{B}(\hat{y})$ denotes the prediction distribution for the entire day, which is not fully accessible until the end of the day. We thus make the below presumption and try to use partial data of the given day for distribution estimation despite a certain delay in model serving.

\subsubsection{Presumption II: $\frac{\mathcal{B}(y)}{\mathcal{B'}(y)}=\frac{\mathcal{B}_h(y)}{\mathcal{B}'_h(y)}$} We denote $\mathcal{B}_h(y)$ as label distribution computed with the data from the midnight to the $h$-th hour of the day. This assumption states that, within one day, the disparity between two sales promotions remains steady no matter how many hours of data are employed\footnote{However, $\mathcal{B}_h(y)$ itself is not stable and may change with $h$. For example, the CVR at night is larger than that in the morning as customers are often at work during the day time and shop at night. We put more discussion in Appendix~\ref{apd:dis_pre}.}. Using the actual data, we study the stability of $\mathcal{B}_h(y) / \mathcal{B}'_h(y)$ with different selections of the hour $h$. As shown in Figure~\ref{fig:cvr-ratio}, we do observe a relatively stable ratio as the largest difference within one day does not exceed 5\%. In the production system, we set $h=10$; therefore, the model serving is delayed by 10 hours. However, we argue that it won't affect much as the conversion mostly happens during the other half of the day, and such a delay is also important for feature computation in historical data retrieval, as mentioned in Section~\ref{subsec:hdrr-vector-repsentation}.

Under this presumption, the loss in Equation~\ref{eq:loss-after-presumption1} now turns into:
\begin{equation}
% \small 
\begin{aligned}
\label{eq:loss-presumption2} 
\mathcal{L}
= \int \frac{\mathcal{B}_h(y)}{\mathcal{B}'_h(y)} \cdot \ell(x,y) \ \mathrm{d} \mathcal{B}'(x,y) 
\end{aligned} 
\end{equation}

Again, $\mathcal{B}'_h(y)$ can be computed from the historical data. In terms of $\mathcal{B}_h(y)$, we repeat the same reasoning process as did in Section~\ref{subsec:presumption1} and acquire the below Equation~\ref{eq:bbse-hour} for $\mathcal{B}_h(\hat{y})$. Now, we have both $\mathcal{B}_h(\hat{y})$ and $\mathcal{B}'_h(\hat{y}|y)$ available at the ($h+1$)-th hour. With $\mathcal{B}_h(y)$ being the only unknown variable, we can solve the equation and obtain an analytic solution. More details about the final solution can be found in Appendix~\ref{apd:solution}.
% \begin{equation}
% \begin{aligned}
% \label{eq:bbse-hour} 
% \mathcal{B}_h(\hat{y}) 
% & = \int \mathcal{B}'_h(\hat{y}|y) \cdot \mathcal{B}_h(y) \ \mathrm{d}y. 
% \end{aligned} 
% \end{equation} 
\begin{equation}
% \small 
\begin{aligned}
\label{eq:bbse-hour} 
\mathcal{B}_h(\hat{y}) 
& = \sum_{y \in \{0, 1\}} \mathcal{B}'_h(\hat{y}|y) \cdot \mathcal{B}_h(y). 
\end{aligned} 
\end{equation}

\subsection{Effective Fine-tuning with TransBlock} 
\label{sec:transblock}
The above two sections haven't discussed the detailed modeling architecture for the fine-tuning module and how it interacts with the main model, which is the focus of this section. However, in practice, we encounter a serious overfitting problem during fine-tuning. This has been studied as the one-epoch issue in prior studies, that is, the model will be overfitted once the same data has been seen twice~\cite{zhang2022towards,zhu2021open,zhou2018deep}. We thus develop a TransBlock module to resolve this challenge. It is worth emphasizing that TransBlock also captures the transition between non-promotional and promotional conversions, making the predictions easily interpretable.

\begin{figure}[t]
    \centering
    \includegraphics[width=\columnwidth]{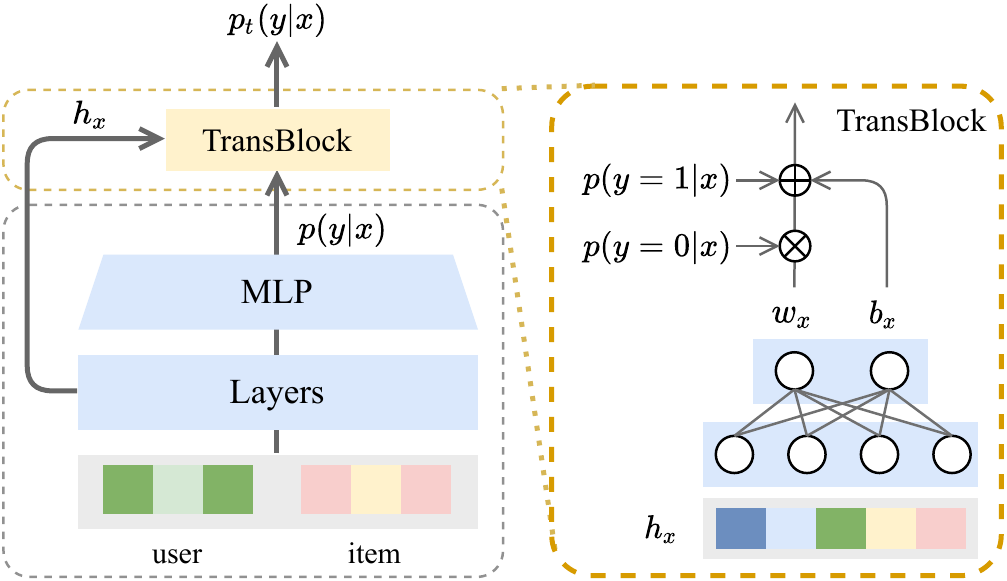} 
    \caption{The overall architecture of TransBlock. To overcome the one-epoch problem, the main model and TransBlock are updated with separate training configurations.} 
    \label{fig:transblock}
    % \vspace{-1em}
\end{figure} 

As illustrated by Figure~\ref{fig:transblock}, TransBlock introduces a few extra parameters stacked on top of the main model. During fine-tuning, the parameters of TransBlock and the main model will be both updated through training the retrieved data $\mathcal{B}'$. 
To prevent the overfitting problem caused by re-training with the same data, we set an extensively smaller learning rate for the main model while keeping the regular learning rate for TransBlock.
% \footnote{In the production deployment, the learning rate of the main model is set to 1e-5, whereas the learning rate of TransBlock is set to 1e-3.}. 
In this way, the main model trained from the non-promotional data will be less impacted, while the TransBlock module helps this main model quickly adapt to the promotion mode after fine-tuning.

To capture the difference between non-promotional conversion and promotional conversion, we explicitly model such a pattern with the below Equation~\ref{eq:transblock-probability}. More specifically, we compute the probability of promotional conversion with the following three components: (1) the probability of non-promotional conversion $p(y=1|x)$ obtained from the main model, (2) the transition from non-promotional non-conversion to promotional conversion $v_x \cdot p(y=0|x)$, and (3) the transition from non-promotional conversion to promotional non-conversion $u_x \cdot p(y=1|x)$, that is,
\begin{equation} 
% \small 
\begin{aligned} 
\label{eq:transblock-probability} 
p_t(y=1|x) = p(y=1|x) + v_x \cdot p(y=0|x) - u_x \cdot p(y=1|x) 
\end{aligned} 
\end{equation} 
Here, $u_x$ and $v_x$ denote the instance-level transition probabilities with regard to the input feature $x$. They are both obtained from TransBlock. With a further expansion of the equation, we acquire: 
\begin{equation}
% \small 
\begin{aligned}
\label{eq:transblock} 
p_t(y=1|x) 
=& p(y=1|x) + v_x \cdot p(y=0|x) - u_x \cdot p(y=1|x) \\ 
=& p(y=1|x) + v_x \cdot p(y=0|x) - u_x \cdot (1-p(y=0|x)) \\
=& p(y=1|x) + (v_x+u_x) \cdot p(y=0|x) - u_x \\
=& p(y=1|x) + \underbrace{w_x \cdot p(y=0|x) + b_x}_{w_x=v_x+u_x, b_x=-u_x}. 
\end{aligned} 
\end{equation} 

In practice, we utilize a simple Multi-layer Perceptron (MLP) that takes the transformed input feature $h_x$ (after the embedding transformation of $x$) and the predictions of the main model as input, and outputs $w_x$ and $b_x$ for computing the probability of promotional conversion. As illustrated by Equation~\ref{eq:transblock-wb}, $W_\text{trans}$ and $b_\text{trans}$ are trainable parameters in TransBlock. 
\begin{equation}
% \small
\begin{aligned}
\label{eq:transblock-wb} 
\begin{bmatrix}
w_x \\ b_x 
\end{bmatrix} = W_\text{trans} \cdot h_x  + b_\text{trans} 
\end{aligned} 
\end{equation} 

Since the output of TransBlock has now turned into the summation of three different components, there is no guarantee of the output range. Hence, we truncate the final prediction between [0,1] with the below Equation~\ref{eq:transblock-truncation}, 
\begin{equation}
% \small 
\begin{aligned}
\label{eq:transblock-truncation}
p'_t(y=1|x) = \left\{
    \begin{matrix}
        0, &  p_t(y=1|x) < 0 \\
        1, &  p_t(y=1|x) > 1 \\
        p_t(y=1|x), &  \text{otherwise}, \\
    \end{matrix}
\right. 
\end{aligned} 
\end{equation}

\subsection{Optimization} 
\label{subsec:optimization}
With the above probability $p'_t(y=1|x)$, we then compute the corresponding \textbf{cross-entropy loss} $\ell(x,y)$. Along with the importance weight $\frac{\mathcal{B}_h(y)}{\mathcal{B'}_h(y)}$ (calculated in Section~\ref{sec:dsc}), we further place them back to Equation~\ref{eq:loss-presumption2} for obtaining the final fine-tuning loss. Up to now, we have a complete picture of the loss function for HDR. During fine-tuning, \textbf{$W_\text{trans}$ and $b_\text{trans}$ are updated with a regular learning rate $\eta_{1}$, whereas the original main model utilizes a much smaller learning rate $\eta_{2}$.} Such an optimization trick is important and significantly improves model performance as shown in Section~\ref{sec:res_tb}. It is worth noting that HDR only concentrates on the fine-tuning of the main model for better adaptation to the promotion mode, i.e., the ${Finetune(\mathcal{B}')}$ term in Equation~\ref{eq:training-framework}. 

\subsection{Overall Online Learning Process for HDR} 
HDR introduces a new training-and-finetuning paradigm that includes an automated historical data retrieval module and a model fine-tuning module. As a consequence, the original online learning process as illustrated by Figure~\ref{fig:odl} is no longer applicable. As a result, we redesign the existing online learning framework by injecting a fine-tuning node. This new online learning process has been validated and successfully deployed in the display advertising platform of Alibaba, serving tens of millions of users on daily basis.

\begin{figure}[t]
    \centering
    \includegraphics[width=1.0\columnwidth]{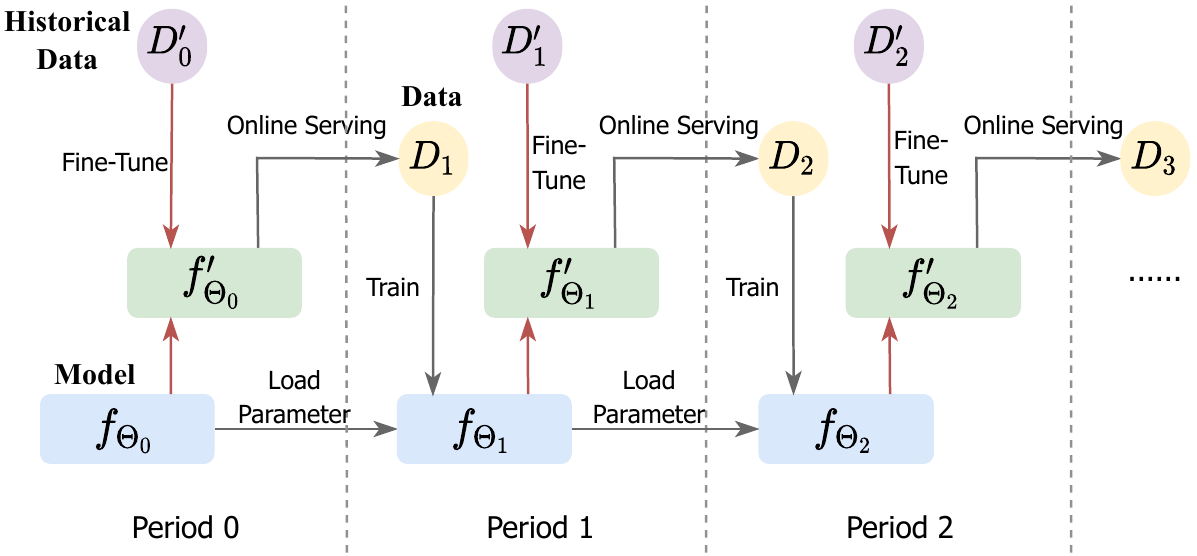} 
    \caption{An illustration of the training-and-finetuning online learning process for the proposed HDR approach.} 
    \label{fig:dual_odl}
\end{figure}

More specifically, as illustrated by Figure~\ref{fig:dual_odl}, the new online learning framework keeps the original process intact, that is the main model $f_\Theta$ still updates on a daily basis. In addition, the fine-tuning node reloads the parameters from the main model and initializes the parameters for TransBlock. 
Once the retrieved data is ready, the fine-tuning process will be invoked automatically. After the fine-tuned model $f'_\Theta$ is ready, it will be deployed to the production system for online serving. Note that, the fine-tuning node also runs on a daily basis to match the update frequency for the main model. 

\section{Experiment Setup} 
\label{sec:exp}
To understand the effectiveness of the proposed HDR approach, we conduct an extensive number of offline and online experiments. This section starts with introducing the dataset for the offline experiments, which is collected from the display advertising system of Alibaba. Afterward, we introduce a set of evaluation metrics to measure the calibration ability and ranking performance of the proposed approach. For a complete understanding of HDR, we further compare its performance with a variety of baselines. 

\subsection{Dataset} 
Due to the lack of relevant public datasets, we first collect the user click log from the online display advertising system of Alibaba, which contains almost half-year data from 2022/05/22 to 2022/09/06. For each click, we store a binary 0/1 label indicating whether such a click leads to an ultimate purchase in three days. In addition, we also record the specific timestamp when the purchase occurred, and `N/A' when there are no purchases. In this dataset, the average number of clicks (purchases) each day is around 40 million (600 thousand). For offline experiments, we treat the click and purchase information on 2022/09/06 (i.e., the promotion time for 9.9 Sales) as the testing data, and the rest as the training data. 

\subsection{Evaluation Metrics}
In this paper, we evaluate both the ranking performance and calibration ability of the proposed HDR approach.

\subsubsection{Ranking Performance} We first regard CVR prediction as a classification problem and thereby utilize \textbf{AUC}~\cite{schutze2008introduction} to evaluate its ranking ability. AUC is commonly adopted to measure how well the model differentiates between the distributions of different classes, and thus it is widely used in the two-class classification problem.

\subsubsection{Calibration Performance} Calibration measures how well a prediction aligns with the actual conversion rate, which is of great importance for applications such as online advertising~\cite{sheng2022joint}. To measure the calibration performance, we employ three widely-adopted metrics including \textbf{LogLoss}, expected calibration error (\textbf{ECE}), and predicted CVR over the actual CVR (\textbf{PCOC})~\cite{yan2022scale,huang2022mbct}.

LogLoss measures the sample-level calibration error, while PCOC (or ECE) offers a macro view of the overall dataset-level (or subset-level) calibration performance. A smaller value of LogLoss or ECE usually implies a better calibration performance; and for PCOC, our goal is to obtain a value that is close to 1~\footnote{We put the computation of these three calibration metrics in Appendix~\ref{apd:metrics}.}.

\subsubsection{Online Performance} With respect to the online A/B testing, we employ a set of top-line business metrics that include conversion rate (CVR), Revenue Per Mille (RPM), and Return Over Investment (ROI). Online CVR is a direct measure of prediction performance, and RPM and ROI are important metrics with regard to the industrial advertising system. RPM reflects the monetization ability of the production system, whereas ROI suggests the actual effect that our system can help improve the business of retailers.

\subsection{Compared Methods} 
\label{subsec:compared-methods}
To understand the effectiveness of the proposed HDR approach, we introduce the below CVR prediction methods for comparison. 

\begin{itemize}[leftmargin=*] 
    \item {\textbf{Base.}} DEFER~\cite{gu2021real} is adopted as the base approach since it is our production CVR model and serves the main traffic of Alibaba display advertising system. Note that this method is one of the State-Of-The-Art delayed feedback models for CVR estimation.

    \item {\textbf{ES-DFM.}} ES-DFM~\cite{yang2021capturing} is another delayed feedback modeling algorithm that optimizes the expectation of true conversion distribution via importance sampling. It estimates the importance weight for each training sample and utilizes the estimated weight to revise the loss function for model training. 

    \item {\textbf{DEFUSE.}} DEFUSE~\cite{chen2022asymptotically} is designed to tackle the delayed feedback problem in CVR prediction as well. Specifically, it divides the data samples into four groups --- immediate positive samples, delayed positive samples, fake negative samples, and real negative samples, and further re-designs the importance sampling strategies for each of them, respectively.
    
    \item {\textbf{Base w/o DFM.}} The irregular conversion pattern during sales promotions violates the assumption of most existing delayed feedback models. For this reason, we experiment with a non-delayed-feedback modeling approach by discarding recent data without the complete conversion labels. Since the attribution window for conversion is three days in the production system, this method discards user behaviors of the recent two days.
     
    \item {\textbf{Base w/ promotion indicator.}} In addition to the fine-tuning of historical promotion data, another viable approach is to differentiate the promotion and non-promotion data with a binary indicator feature. In this way, the model may learn separate patterns for promotional and non-promotional conversions. 
    
    \item {\textbf{Base w/ direct retraining.}} The simplest data reuse approach is to directly re-train the production model on top of the retrieved historical data. 
    In fact, such a method is equivalent to HDR with the removal of the distribution shift correction module and the TransBlock module. 
    However, this may introduce the one-epoch problem as mentioned in ~\citet{zhang2022towards}.
    
    \item {\textbf{HDR.}} To further analyze the usefulness of each module in HDR, we conduct a set of ablation studies on the distribution shift correction module and the TransBlock module, in which we take into account \textbf{HDR w/o DSC} (distribution shift correction) and \textbf{HDR w/o TransBlock}, respectively. 
\end{itemize}

It is worth noting that \textbf{ES-DFM}, \textbf{DEFUSE}, and \textbf{Base w/o DFM} are introduced for the purpose of understanding the effectiveness of different delayed feedback models, particularly how well they perform during sales promotions. 
The introduction of \textbf{Base w/ promotion indicator} and \textbf{Base w/ direct retraining} is to examine the usefulness of data utilization methods other than HDR. Moreover, by examining a number of HDR variants, we can acquire a better understanding of the efficacy for each HDR component\footnote{We put the the details of implementation in Appendix~\ref{apd:impl}.}.

\section{Experimental Results} 
This section starts with analyzing model performance for all of the compared methods in Section~\ref{subsec:compared-methods}. After that, we conduct a number of ablation studies to understand the effectiveness of the distribution shift correction algorithm, the TransBlock module, and the automated historical data retrieval process. In the end, we evaluate the online performance of HDR after production deployment.

\subsection{Overall Performance} 
Table~\ref{tab:result_offline} provides a comparison of the ranking ability and calibration performance for all of the methods listed in Section~\ref{subsec:compared-methods}. We find that the \textbf{Base} model performs poorly on the promotion data, in which the AUC metric is 0.793 and PCOC is only 0.471. This is in line with the declined online performance as illustrated by Figure~\ref{fig:metrics-in-promotions} and, again, demonstrates the necessity to develop a promotion-compatible CVR prediction model for production use.

We further investigate the effectiveness of different delayed feedback modeling techniques. According to Table~\ref{tab:result_offline}, we identify that same as DEFER (i.e., the \textbf{Base} model), other delayed feedback models including \textbf{ES-DFM} and \textbf{DEFUSE} are still unable to alleviate the deteriorated model performance. On the contrary, after the removal of the delayed feedback module (i.e., \textbf{Base w/o DFM}), the Base model now outperforms \textbf{ES-DFM} and \textbf{DEFUSE}. Both the ranking and calibration metrics have increased by a large proportion (despite the PCOC metric remaining at a low level). This illustrates the failure of the underlying assumption for delayed feedback models during sales promotions and calls for the development of novel techniques to deal with the fluctuation of conversion behaviors.

In considering the uniqueness of the promotional conversion patterns, \textbf{HDR} takes a data reuse and fine-tuning approach. However, there might also be other simpler ways for promotional data utilization. Here, we evaluate the usefulness of two additional methods as mentioned in Section~\ref{subsec:compared-methods}, which are \textbf{Base w/ promotion indicator} and \textbf{Base w/ direct retraining}. Again, from Table~\ref{tab:result_offline}, we find that both of them produce on-par performances as the \textbf{Base} model. The introduction of a binary promotion indicator does not help much, which might be due to the forgetting of previous promotional conversion patterns in continuous online learning. In the meantime, direct retraining of the historical data also fails to achieve a good performance. In fact, the AUC even drops by 2.8\%. This is due to the one-epoch overfitting problem as discussed in ~\citet{zhang2022towards}. Overall, this again demonstrates the necessity to develop more proper historical data reuse mechanisms. 

\begin{table}
\centering
% \small 
% \renewcommand{\arraystretch}{0.9} 
\caption{A comparison of AUC, PCOC, Logloss, and ECE performance for different experiments.} 
\label{tab:result_offline}
\resizebox{0.95\columnwidth}{!}{
\begin{tabular}{lcccc}
\toprule 
% Methods & AUC $\uparrow$ & PCOC & Logloss $\downarrow$ & ECE $\downarrow$ \\ 
Methods & AUC & PCOC & Logloss & ECE \\ 
\midrule 
Base                  & $0.793$ & $0.471$ & $0.143$ & $0.0212$ \\ 
ES-DFM                & $0.798$ & $0.441$ & $0.149$ & $0.0245$ \\ 
DEFUSE                & $0.806$ & $0.452$ & $0.145$ & $0.0151$ \\ 
Base\small{ w/o DFM}  & $0.826$ & $0.539$ & $0.103$ & $0.0120$ \\ 
\midrule 
Base\small{ w/ promotion indicator}  & $0.801$ & $0.479$ & $0.147$ & $0.0191$ \\ 
Base\small{ w/ direct retraining}    & $0.779$ & $0.718$ & $0.156$ & $0.0241$ \\ 
\midrule 
{HDR}~                          & $\textbf{0.875}$ & $\textbf{0.990}$ & $\textbf{0.090}$ & $\textbf{0.0010}$ \\ 
\small{\quad w/o DSC}~                    & $0.874$ & $0.803$ & $0.091$ & $0.0051$ \\ 
\small{\quad w/o TransBlock}~             & $0.771$ & $0.861$ & $0.167$ & $0.0232$ \\
\bottomrule 
\end{tabular}} 
\end{table} 

Our proposed \textbf{HDR} approach actually provides such a mechanism. As reported in Table~\ref{tab:result_offline}, it outperforms all of the baseline methods in terms of both ranking ability and calibration performance. Specifically, HDR outperforms the Base method by 10.3\% on AUC, and the PCOC is 0.99 which almost reaches the ideal value of 1.0. Meanwhile, it also achieves the best performance for Logloss and ECE. They all demonstrate the effectiveness of HDR. 

\subsection{Ablation Study} 
To further examine the usefulness of each module in HDR, we conduct the below ablation studies in this section. 

\begin{table} 
\centering 
% \small 
% \renewcommand{\arraystretch}{0.9} 
\caption{Model performance over different values of the learning rate $\eta_{2}$. Hereafter, $\eta_{2}$ is set to $1e^{-5}$ for production models.}
\label{tab:result_transblock} 
\resizebox{0.7\columnwidth}{!}{
\begin{tabular}{lcccc} 
\toprule 
% Setting & AUC $\uparrow$ & PCOC & Logloss $\downarrow$ & ECE $\downarrow$ \\ 
Setting & AUC & PCOC & Logloss & ECE \\ 
\midrule 
$\eta_{2}=0$       & $0.866$ & $0.953$ & $0.094$ & $0.0015$ \\ 
$\eta_{2}=1e^{-5}$ & $0.875$ & $0.990$ & $0.090$ & $0.0010$ \\ 
$\eta_{2}=1e^{-4}$ & $0.842$ & $1.012$ & $0.085$ & $0.0023$ \\ 
\bottomrule 
\end{tabular}} 
\end{table} 

\subsubsection{The Effect of Distribution Shift Correction (DSC)} 
To understand the influence of DSC, we experiment with the \textbf{HDR w/o DSC} approach by removing the DSC module. This is achieved by setting $\mathcal{B}_h(y) / \mathcal{B}'_h(y)$ to 1.0 in Equation~\ref{eq:loss-presumption2}. As shown in Table~\ref{tab:result_offline}, we acquire a roughly identical AUC score as the \textbf{HDR} model. However, this leads to a significant drop in calibration performance -- there is a nearly 20\% decline in PCOC, and the ECE also increases by a large extent (since ECE measures the error, a larger value indicates worse performance). This illustrates the importance of the DSC module in improving calibration performance.

\subsubsection{The Effect of TransBlock} 
\label{sec:res_tb}
We also experiment with the \textbf{HDR w/o TransBlock} approach by removing the TransBlock module. More concretely, this method still utilizes the retrieved historical data for model fine-tuning, and re-weights the loss function through DSC, whereas the purposefully-designed TransBlock probability in Equation~\ref{eq:transblock-truncation} is discarded for loss computation. As reported in Table~\ref{tab:result_offline}, we find a significant drop in model performance compared to HDR, which illustrates the usefulness of TransBlock once more.

As mentioned in Section~\ref{subsec:optimization}, during fine-tuning, we adopt two separate learning rates: $\eta_1$ for TransBlock and $\eta_2$ for the original main model. In this section, we set $\eta_1$ to $1e^{-3}$, and examine the model performance over different values of $\eta_2$. In Table~\ref{tab:result_transblock}, we observe a sub-optimal model performance when $\eta_2$ is not small enough. This is due to the one-epoch issue as mentioned in ~\citet{zhang2022towards}. However, a complete freeze of the original model parameters (i.e., $\eta_2=0$) is not the best choice. Thus, the selection of a proper learning rate is another important step for fine-tuning.

\begin{table} 
\centering 
\caption{Case studies for the historical data retrieval results.} 
\label{tab:retrieval_result} 
\resizebox{\columnwidth}{!}{
\begin{tabular}{c|cc|c|c} 
\toprule[0.9pt]
 & \multicolumn{2}{c|}{Date} &  CVR & Similarity \\ 
\midrule 
Target                         & 2022/09/06 & 9.9 Sales & \textbf{x}      & - \\ 
\midrule
\multirow{2}{*}{Top Two Dates} & 2022/08/08 & 8.8 Sales   & \poscl 0.82 \textbf{x} & \poscl 0.9923 \\ 
                               & 2022/06/14 & 6.18 Sales & \poscl 0.84 \textbf{x} & \poscl 0.9919 \\ 
Randomly-picked Date           & 2022/05/28 & No Sales   & \negcl 0.58 \textbf{x} & \negcl 0.9212 \\ 
\midrule[0.9pt] 
Target                         & 2022/08/08 & 8.8 Sales & \textbf{y}      & - \\ 
\midrule
\multirow{2}{*}{Top Two Dates} & 2022/07/12 & Summer Sales & \poscl 0.97 \textbf{y} & \poscl 0.9937 \\ 
                               & 2022/07/31 & Qixi Sales   & \poscl 0.83 \textbf{y} & \poscl 0.9910 \\ 
Randomly-picked Date           & 2022/05/28 & No Sales     & \negcl 0.71 \textbf{y} & \negcl 0.8902 \\ 
\bottomrule[0.9pt]
\end{tabular}}
 \begin{tablenotes}
    \footnotesize 
    \item[*] * For the purpose of data anonymity, we only report the relative CVR values. 
\end{tablenotes}
\end{table} 

\subsubsection{The Effect of Historical Data Retrieval} 
As shown in Table~\ref{tab:retrieval_result}, we provide several case studies to better illustrate the accuracy of historical data retrieval as well. The first one is to find similar promotions for 9.9 Sales (on the date of 2022/09/06 and with a CVR of \textbf{x}). The top two dates we retrieved are 8.8 Sales on 2022/08/08 with a CVR of 0.82\textbf{x}, and 6.18 Sales on 2022/06/14 with a CVR of 0.84\textbf{x}. The second one is to seek similar promotions for 8.8 Sales (on the date of 2022/08/08 and with a CVR of \textbf{y}). The top two retrieved results are Summer Sales on 2022/07/12 with a CVR of 0.97\textbf{y}, and Qixi Sales on 2022/07/31 with a CVR of 0.83\textbf{y}. In addition to the top two results, we also randomly pick a date with a much lower similarity score. Clearly, the overall CVR of such a random date differs from the target promotion date by a very large extent. This proves the effectiveness of the data retrieval process in HDR.

% \subsubsection{\chan{Concern}}  
% Here, we discuss an underlying concern of HDR in production systems. Practical advertising systems is a non-stationary as advertisement campaigns (i.e., candidates) are constantly changing~\cite{chapelle2014modeling}. Especially during the big promotion cycle, the surge in marketing demands brings lots of new advertisement campaigns. The models fine-tuned on outdated historical data may not perform as well in such a complex system with old and new campaigns. To answer this concern, it is worth introducing an online experimental environment by developing HDR in our production systems in Section~\ref{sec:answer}.

\subsection{Production Deployment} 
HDR has been deployed on the display advertising system of Alibaba, serving the main traffic during the Double 11 Sales in 2022. Here, we provide the details of our production deployment. 

\subsubsection{System Deployment}
In the historical data retrieval stage, we create a daily map-reduce task to build a vector representation for each day. After that, we compute the cosine similarity between the present day and each of the historical dates based on their vector representations and then select the top two dates for fine-tuning. Since we also adopt several features computed from  user behaviors in the first 10 hours of the present day, this stage usually completes around 10:30 am each day.

During fine-tuning, we need to calculate the importance weight $\frac{\mathcal{B}_h(y)}{\mathcal{B}'_h(y)}$ for the loss function in Equation~\ref{eq:loss-presumption2}. In practice, $\mathcal{B}'_h(y)$ is estimated from the historical data, and $\mathcal{B}_h(y)$ can be calculated with Equation~\ref{eq:bbse-hour}. As for $\mathcal{B}_h(y)$, we need to access the data for both $\mathcal{B}_h(\hat{y})$ and $\mathcal{B}'_h(\hat{y}|y)$. The former is obtained through parsing the real-time prediction log, and the latter is collected by running a batch prediction script to score the historical data with the same prediction model. Note that model fine-tuning is conducted after the data retrieval stage; thereby, the fine-tuned model is usually accessible for online serving at around 11:00 am each day. 

\subsubsection{Ranking and Calibration Performance} 
\label{sec:rcm}
\begin{figure}[t]
    \centering
    \includegraphics[width=0.95\columnwidth]{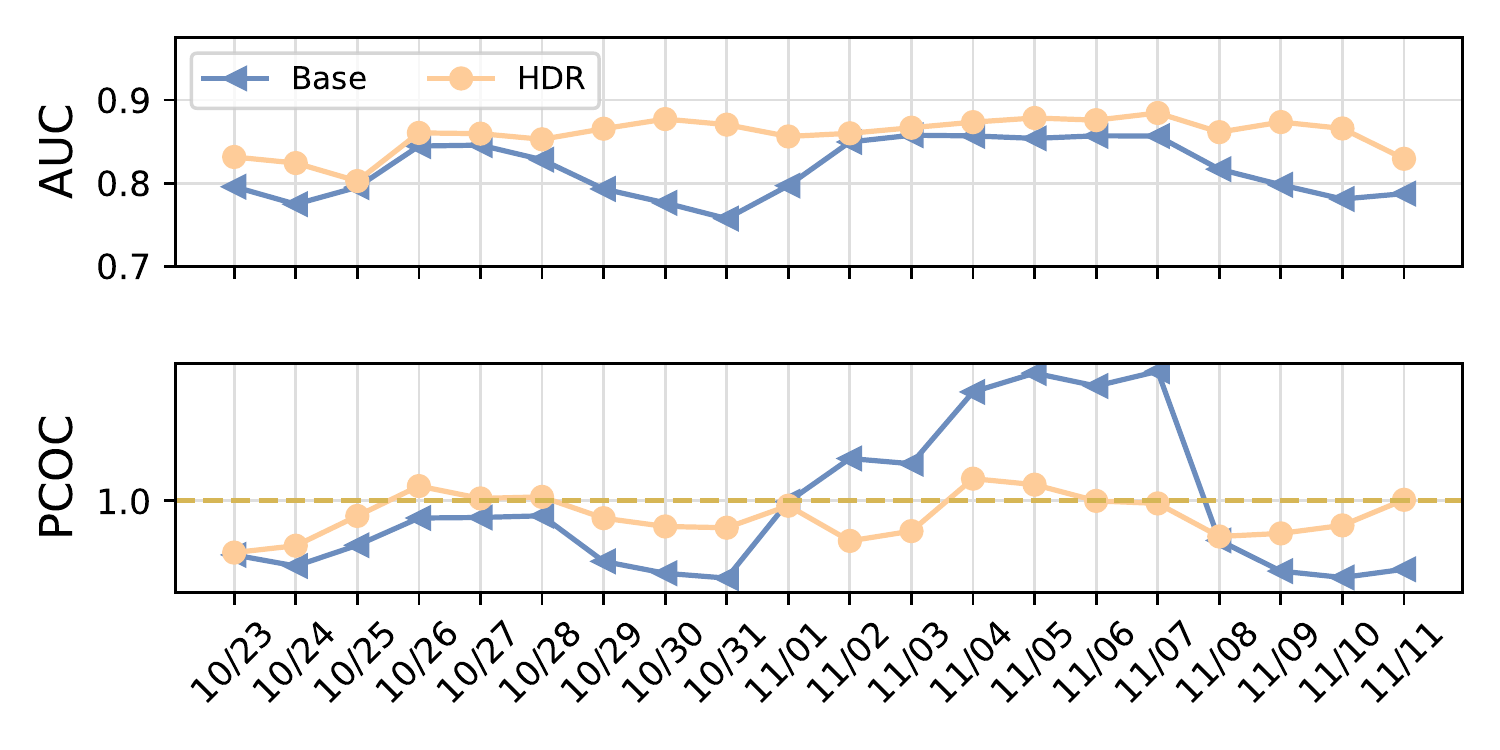} 
    \caption{A comparison of the model performance (AUC and PCOC) between Base and HDR with online A/B testing.} 
    \label{fig:online_metric}
\end{figure}

We first compare the model performance between HDR and Base through online A/B testing. Here, we adopt both the AUC and PCOC metrics to measure the ranking and calibration performance, respectively. As shown in Figure~\ref{fig:online_metric}, throughout the whole sales promotion, HDR exhibits consistently superior performance over the Base model in terms of both AUC and PCOC. More importantly, it nearly resolves the problem of substantial performance drop around the promotion dates (e.g. 10/24, 10/31, and 11/10). With the proposed HDR approach, the PCOC and AUC curves have become much flatter.

To further understand the model performance in different product categories, we compute the corresponding PCOC metrics. According to Table~\ref{tab:online_pcoc}, we observe severe underestimation of the Base model in the listed categories, in which the predicted CVR is only 1/7 of the actual CVR. HDR effectively tackles this problem by utilizing the historical promotion data, and PCOC has largely improved to 0.7. Note that the CVR prediction of sports shoes and sportswear, which are seriously underestimated by the Base model, now performs as well as other product categories. This proves that our method is capable of capturing the category difference.

\subsubsection{Production Metrics} 
HDR has been deployed in the display advertising system of Alibaba. During the Double 11 Sales of 2022, we conduct a strict online A/B test to validate the effectiveness of HDR. 
% As shown in Table ~\ref{tab:online}, c
\textbf{Compared to the production baseline, HDR achieves a 9\% increase in RPM}. Such a significant improvement mainly comes from the optimization of the calibration ability, which allows the online advertising system to measure the value of production traffic in a more accurate way. \textbf{In the meantime, there is a 16\% increase in CVR and an 11\% increase in ROI}, which are largely ascribed to the improved ranking ability (i.e. AUC). 
% Such improvements are really significant to our business. 

 % \begin{table}
 % \centering
 % \small 
 % \caption{A comparison of the online performance between HDR and Base during the 11.11 Sales in 2022.}
 % \label{tab:online}
 % \begin{tabular}{lccc}
 %     \addlinespace
 %     \toprule
 %     Metric & RPM & CVR & ROI \\
 %     \midrule
 %     Lift Rate & $+9\%$ & $+16\%$ & $+11\%$ \\
 %     \bottomrule
 % \end{tabular}
 % \label{tab:ol_performance}
 % \end{table}

 \begin{table} 
\centering 
% \tiny
\caption{An comparison of the PCOC metrics between Base and HDR over different product categories on 2022/10/31.} 
\label{tab:online_pcoc} 
\resizebox{0.525\columnwidth}{!}{
\begin{tabular}{l|cc} 
\toprule 
    % \multirow{2}{*}{Category} & \multicolumn{2}{c}{PCOC} \\ \cmidrule{2-3} 
    Category &  Base  &  HDR \\ 
    \midrule 
    women's clothing      & 0.139 & 0.695 \\ 
    % men's clothing        & 0.176 & 0.677 \\ 
    children's clothing   & 0.135 & 0.721 \\ 
    sports shoes          & \poscl 0.054 & \poscl 0.735 \\ 
    women's shoes         & 0.152 & 0.699 \\ 
    sportswear            & \poscl 0.071 & \poscl 0.743 \\ 
\bottomrule 
\end{tabular}}
\vspace{-1mm}
\end{table}

\subsubsection{Discussion} 
One potential drawback for HDR is that the reuse of historical data may encounter the problem of data obsolescence. Specifically, online advertising systems are dynamic as active advertisers and advertising campaigns are constantly changing; thus, the active advertisements may differ significantly across different promotions~\cite{chapelle2014modeling}. Particularly, during promotion, the surge of marketing demands also brings lots of new campaigns that are impossible to observe from history. Models fine-tuned on outdated historical data may be sub-optimal or even harmful to online performance. 

However, in practice, we do not observe any sign of counterproductive performance. This might be attributed to the design of the fine-tuning mechanism, in which the original model (with a sparse ID denoting each specific ads campaign) is rarely updated because of the small learning rate, whereas TransBlock (only with a few parameters) learns coarse-grained knowledge, such as category-level difference, for data adaptation. Such a phenomenon is worth further exploration for a better understanding of the historical data reuse mechanism, and we will take it into account in the future.  

% Here, we first discuss some hands-on experiences from experiments and production development. In this research, all experiments are conducted on top of the online display advertising system of Alibaba, in which a well-calibrated estimation of CVR is of great importance. Other machine learning systems such as search and recommendation may not require the prediction score to be exactly calibrated; however, we argue that such an approach is still transferable as it significantly improves the ranking ability as well. On the other hand, without the constraint of calibrated prediction, the model updating frequency can be largely improved by removing the distribution shift correction module. In addition, the one-epoch phenomenon may not exist in machine learning systems that do not utilize deep learning models or adopt much simpler modeling architecture. In this case, they may not need a TransBlock module to address the overfitting problem.

\section{Conclusion} 
In this work, we propose a novel data reuse mechanism, named HDR, to tackle the complicated CVR prediction problem during sales promotions. HDR consists of three components: an automated data retrieval module that seeks the most similar data from historical promotions, a distribution shift correction module that re-weights the retrieved data for better aligning with the target promotion, and a TransBlock module that quickly fine-tunes the original model for better adaptation to the promotion mode. Note that HDR only concentrates on fine-tuning the production model with historical promotion data whereas the original online learning process keeps intact. 
Experiments conducted on top of real-world datasets demonstrate the effectiveness of HDR, as it improves both ranking and calibration metrics to a large extent. 
% Experiments prove that HDR could improves both ranking and calibration metrics to a large extent. 
HDR has also been deployed on the display advertising system in Alibaba, bringing a lift of $9\%$ RPM and $16\%$ CVR during Double 11 Sales in 2022. 

In light of the great performance of HDR, we would like to explore other occasions that are also suitable for data reuse. For instance, we identify that conversion also differs between weekdays and weekends, middle-of-month and end-of-month, etc. Moreover, we believe that the data reuse mechanism is not limited to the historical data, we can also explore utilizing the data from other scenes. In the end, we would like to develop a comprehensive data-centric mechanism with a unified data utilization technique.

\section*{Acknowledgements} 
We would like to thank the anonymous reviewers for their constructive comments. 
This work was supported by Alibaba Group through Alibaba Research Intern Program, the National Natural Science Foundation of China (No.U2133218), the National Key Research and Development Program of China (No.2018YFB0204304) and the Fundamental Research Funds for the Central Universities of China (No.FRF-MP-19-007 and No. FRF-TP-20-065A1Z). 

\balance 
\bibliographystyle{ACM-Reference-Format}
\bibliography{chapters/reference}

\clearpage 
\appendix

\begin{table}[t]
\centering 
\small 
% \caption{An analysis of the conditional input distribution $p(x|y)$ for 9.9 Sales (in 2022/09) and 3.8 Sales (in 2023/03). The symbol keep the same meaning as in Table~\ref{tab:label-shift}.} 
\caption{An analysis of the conditional input distribution $p(x|y)$ for 9.9 Sales (in 2022/09) and 3.8 Sales (in 2023/03).} 
\label{tab:label-shift-appendix}
\resizebox{\columnwidth}{!}{
\begin{tabular}{c|cc|cc|cc} 
\toprule 
\multirow{2}{*}{User Group $\times$ Category} & \multicolumn{2}{c}{9.9 Sales} & \multicolumn{2}{|c}{3.8 Sales} & \multicolumn{2}{|c}{No Sales} \\ \cmidrule(lr){2-7}
& $y=0$ & $y=1$ & $y=0$ & $y=1$ & $y=0$ & $y=1$ \\ 
\midrule 
$\text{U}_{1}, \text{C}_{1}$ & 1.1013\% & 0.3595\% & 1.1524\% & 0.3573\% & 0.0235\% & 0.0938\% \\ 
$\text{U}_{1}, \text{C}_{2}$ & 0.0065\% & 0.0575\% & 0.0068\% & 0.0579\% & 0.0014\% & 0.1309\% \\ 
\midrule 
$\text{U}_{2}, \text{C}_{1}$ & 0.3441\% & 0.1597\% & 0.3861\% & 0.1601\% & 0.0039\% & 0.1109\% \\ 
$\text{U}_{2}, \text{C}_{2}$ & 0.1738\% & 0.0288\% & 0.1922\% & 0.0298\% & 0.0098\% & 0.1405\% \\ 
\bottomrule 
\end{tabular}
}
\vspace{-3mm}
\end{table} 

\section{Solution for $\mathcal{B}_h(y)$} 
\label{apd:solution}
For a short representation, we rewrite Equation~\ref{eq:bbse-hour} in the matrix form as shown in Equation~\ref{eq:bbse-hour-matrix}. 
\begin{equation}
\begin{aligned}
\label{eq:bbse-hour-matrix} 
\underbrace{\mathcal{B}_h(\hat{y})}_{\textbf{M}_{\hat{y}}}
& = \underbrace{\begin{bmatrix}
 \mathcal{B'}_h(\hat{y}|y=1) & \mathcal{B}'_h(\hat{y}|y=0) 
\end{bmatrix}}_{\textbf{M}'_{\hat{y}|y}}
\cdot 
\underbrace{\begin{bmatrix}
 \mathcal{B}_h(y=1) \\
 \mathcal{B}_h(y=0)
\end{bmatrix}}_{\textbf{M}_{y}} \\
% & \Longrightarrow \textbf{D}_{\hat{y}} = \textbf{D'}_{\hat{y}|y} \cdot \textbf{D}_{y}. 
\end{aligned} 
\end{equation} 

Since $\textbf{M}_{y}$, i.e. $\mathcal{B}_h(y)$, is the only unknown variable, we can reconstruct an optimization problem as illustrated by Equation~\ref{eq:by-optimization-problem} and seek the solution for $\mathcal{B}_h(y)$. Here, the first term (13.1) corresponds to the solution for Equation~\ref{eq:bbse-hour-matrix}, and the second term (13.2) introduces a regularization constraint to keep the stability of result prediction. More specifically, we denote $\mathcal{B'}_h(y)$ as $\textbf{M}'_{y}$ and constrain the estimated $\mathcal{B}_h(y)$ is close to historical $\mathcal{B'}_h(y)$, i.e., $\frac{\mathcal{B}_h(y)}{\mathcal{B'}_h(y)} \rightarrow 1$. In this way, we could regulate the influence of importance weight on the fine-tuning process. 

\begin{equation}
\begin{aligned}
\label{eq:by-optimization-problem} 
\mathop{\arg\min}_{\textbf{M}_{y}} \underbrace{|| \textbf{M}'_{\hat{y}|y} \cdot \textbf{M}_{y} - \textbf{M}_{\hat{y}} ||^2}_\text{(13.1)} + \underbrace{\lambda \cdot || \textbf{M}'_{y} - \textbf{M}_{y} ||^2 }_\text{(13.2)} 
\end{aligned} 
\end{equation} 

The closed-form solution for $\mathcal{B}_h(y)$ is provided in Equation~\ref{eq:y}. Placing it back to Equation~\ref{eq:loss-presumption2}, we acquire our final loss function. 
\begin{equation}
\begin{aligned}
\label{eq:y} 
\textbf{M}_y = \Bigl(
{\textbf{M}'_{\hat{y}|y}}^T\textbf{M}'_{\hat{y}|y} + \lambda \mathbf{I}
\Bigr)^{-1}
\Bigl({\textbf{M}'_{\hat{y}|y}}^T\textbf{M}_{\hat{y}} + \lambda \textbf{M}'_{y}\Bigr)
\end{aligned} 
\end{equation}

\section{Computation of Calibration Metrics}
\label{apd:metrics}
Let $\hat{p}_i$ denote the predicted conversion rate for the $i$-th sample with a conversion label $y_i$, and $N$ stands for the total number of data samples. With these notations, the computation of LogLoss and PCOC can be expressed with the below Equations. In terms of ECE, we first sort data samples by their predicted probabilities and divide them into $K$ buckets, each containing approximately the same number of samples. Then, we compute the prediction error for each bucket and further aggregate the errors of all of the buckets. In the below Equation, $\vmathbb{1}(\hat{p}_i\in{B_k})$ refers to a binary indicator with the value of 1 if the predicted probability locates in the $k$-th bucket $B_ K$, and 0 otherwise.
\begin{equation}
\begin{aligned}
\label{eq:eval-metrics}
\text{LogLoss} &= -\frac{1}{N}\sum_{i=1}^{N} \big(y_i\log\hat{p}_i + (1 - y_i)\log(1-\hat{p}_i)\big); \\ 
\text{PCOC} = &\frac{\sum_{i=1}^{N} \hat{p}_i}{\sum_{i=1}^{N} y_i}; \ \  
\text{ECE} = \frac{1}{N}\sum_{k=1}^{K}|\sum_{i=1}^{N}(y_i-\hat{p}_i)\:\vmathbb{1}(\hat{p}_i\in{B_k})|
\end{aligned} 
\end{equation}

\section{Implementation Details} 
\label{apd:impl}
In practice, we keep the original online learning process intact while only focusing on the model fine-tuning. More specifically, we adopt a single-layer fully-connected neural network for TransBlock\footnote{We provide a demo in \url{https://github.com/MingMTC/HDR}.}, and the number of hidden units is set to [100, 2]. During fine-tuning, we adopt the Adam optimizer~\cite{kingma2014adam} with a learning rate of 1e-3 and a batch size of 5,000 for TransBlock; parameters of the main model are also updated with the Adam optimizer but using a much smaller learning rate of 1e-5. In the meantime, the regularization strength $\lambda$ in Equation~\ref{eq:y} is set to 1.0. 
All of our experimental models are trained with the XDL platform~\cite{jiang2019xdl,zhang2022picasso}.

\section{Illustration for Vector-Based Representation in Retrieval} 
We give an example to illustrate our feature engineering effort. Suppose that a click happens on 2022/09/03, we can then compute the one-day conversion with the purchase data collected in [09/03], the two-day conversion with the data collected in [09/03, 09/04], and so forth. To retrieve similar historical promotion data for 2022/09/06, we utilize the below features in this paper: 

% \begin{itemize}[leftmargin=*] 
% \item One-day CVR for clicks happen on 2022/09/05;
% \item One-day CVR and Two-day CVR for clicks happen on 2022/09/04;
% \item One-day CVR, Two-day CVR and Three-day CVR for clicks happen on 2022/09/03;
% \item The impression ratios of representative categories (such as Toddler, Cosmetics and etc) from 2022/09/03 to 09/05, respectively;
% \item The impression ratios of representative categories in the first 10 hours in 2022/09/06.
% \end{itemize} 

\textbf{(1).} One-day CVR for clicks happen on 2022/09/05;
\textbf{(2).} One-day CVR and Two-day CVR for clicks happen on 2022/09/04;
\textbf{(3).} One-day CVR, Two-day CVR and Three-day CVR for clicks happen on 2022/09/03;
\textbf{(4).} The impression ratios of representative categories (such as Toddler, Cosmetics and etc) from 2022/09/03 to 09/05, respectively;
\textbf{(5).} The impression ratios of representative categories in the first 10 hours in 2022/09/06. 

We utilize all of the above features to build a vector-based representation and retrieve historically similar promotion data for the target promotion date 2022/09/06.

\section{More Analysis about the presumptions} 
\label{apd:dis_pre}
It is notable that the two presumptions are derived from large-scale data analysis and we find that they hold in almost every promotion in our system. 
Here, We provide more analysis to improve the credibility of our presumptions. 
In Table~\ref{tab:label-shift-appendix}, we provide more statistics from the most recent 3.8 Sales (in 2023/03) as well as the 9.9 Sales from the last year. We can see that Presumption I holds for both of them, e.g. p($\text{U}_1$,$\text{C}_1$|y=0) in 9.9 Sales and 3.8 Sales are both around 1.1\%. Moreover, on a normal day without any sales events (No Sales), such a probability has a large deviation from 1.1\%.

As for Presumption II, we also conduct a similar large-scale analysis and do observe a common pattern as illustrated by Figure~\ref{fig:cvr-ratio}, in almost every sales promotion. The rhythm of sales promotions in our system is generally aligned, thereby, we propose Presumption II to simplify our problem. When it is not valid, we have the following alternatives. In practice, we observe that user behaviors are highly correlated to the time difference between the current time and the promotion start time. Therefore, we can calculate $B_h(y)/B'_h(y)$ by aligning the time interval instead of aligning it to an absolute time point. Another alternative is to increase model update frequency. For example, we may calculate $B_h(y)/B'_h(y)$ on an hourly basis, which allows us to capture changes more frequently and accurately.

\end{document}